\documentclass[12pt]{iopart}
\def\0{\mbox{\tiny $0$}}
\def\1{\mbox{\tiny $1$}}
\def\2{\mbox{\tiny $2$}}
\def\3{\mbox{\tiny $3$}}
\def\4{\mbox{\tiny $4$}}
\def\5{\mbox{\tiny $5$}}
\def\6{\mbox{\tiny $6$}}
\def\7{\mbox{\tiny $7$}}
\def\8{\mbox{\tiny $8$}}
\def\9{\mbox{\tiny $9$}}

\def\k{\mbox{\tiny $k$}}
\def\kk{\mbox{\small $k$}}

\def\foh{\mbox{\tiny $\frac{1}{2}$}}
\def\f14{\mbox{\tiny $\frac{1}{4}$}}

\def\s{\mbox{\tiny $s$}}

\def\B{\mbox{\tiny $B$}}

\def\mi{\mbox{\tiny $-$}}
\def\ig{\mbox{\tiny $=$}}
\def\pl{\mbox{\tiny $+$}}

\def\bb#1{\mbox{\footnotesize $(#1)$}}

\begin{document}

\title[Chiral Oscillations and Spin-Flipping]{Dynamics of chiral oscillations - A comparative analysis with spin-flipping}

\author{A. E. Bernardini\dag\ 
\footnote[3]{alexeb@ifi.unicamp.br}
}

\address{\dag\ Instituto de Física Gleb Wataghin, UNICAMP,\\
PO Box 6165, SP 13083-970, Campinas, SP, Brazil}

\date{\today}

\begin{abstract}
Chiral oscillation as well as spin flipping effects correspond to quantum phenomena of fundamental importance in the context of particle physics and, in particular, of neutrino physics.
From the point of view of first quantized theories, we are specifically interested in appointing the differences between chirality and helicity by obtaining their dynamic equations for a {\em fermionic} Dirac-{\em type} particle (neutrino).
We also identify both effects when the non-minimal coupling with an external (electro)magnetic field in the neutrino interacting Lagrangian is taken into account.
We demonstrate that, however, there is no constraint between chiral oscillations, when it takes place in vacuum, and the process of spin flipping related to the helicity quantum number, which does not take place in vacuum.
To conclude, we show that the origin of chiral oscillations (in vacuum) can be interpreted as position very rapid oscillation projections onto the longitudinal direction of momentum.
\end{abstract}




\maketitle 

\section{\label{sec1}Introduction}

Over recent years, the properties of neutrinos \cite{Zub98,Sch03,Alb03} have become the subject 
of an increasing number of theoretical works upon
quantum oscillation phenomena \cite{Zra98,Beu03}.
The flavor mixing models \cite{Nir03,Gou05}, the quantum field prescriptions \cite{Bla95,Giu02,Bla03} and, generically,
the quantum mechanics of oscillation phenomena formalism \cite{Beu03,Vog04,Giu98,Giu02,Ber05A}
have been extensively studied.
Notwithstanding the exceptional ferment in this field, the numerous conceptual
difficulties hidden in the quantum oscillation phenomena represent an intriguing challenge for physicists.
In the last three years,
compelling experimental evidences have continuously ratified
that neutrinos undergo flavor oscillations in vacuum or in matter.
For instance, we refer to the outstanding   
results of the Super-Kamiokande atmospheric neutrino experiment \cite{Fuk02},
in which a significant up-down asymmetry of the high-energy muon events was observed,
to the results of the SNO solar neutrino experiment \cite{Ahm02,Ban03}, in which a direct evidence for the transition of the solar electron neutrinos into other flavors was obtained,
and also to the results of the KamLAND experiment \cite{Egu02} in confirming that the disappearance of solar electron neutrinos is mainly due to neutrino oscillations and not to other types of neutrino conversions \cite{Guz02,Bar02B}.
The experimental data could be completely interpreted and understood in terms
of three flavor quantum numbers, excepting by the LSND anomaly \cite{Ana98,Agu01,Ban03}
which permit us to speculate the existence of (at least) a fourth neutrino (flavor?) which
has to be inert.
In fact, the hypothesis of mixing between known neutrino species (electron, muon and tau)
and higher mass neutrinos including sterile neutrinos was naively studied in the literature \cite{Gel79}.
The neutrino spin-flipping attributed to some dynamic external \cite{Oli90}
interacting process, which comes from the non-minimal coupling of a magnetic moment with an external electromagnetic
field \cite{Vol81}, was formerly supposed to be a relevant effect in the context of the solar-neutrino puzzle. 
As a consequence of a non-vanishing magnetic moment interacting with an
external electromagnetic field, left-handed neutrinos could change their helicity (to right-handed)\cite{Bar96}.
The effects on flavor oscillations due to external magnetic interactions in a kind of
chirality-preserving phenomenon were also studied \cite{Oli96} but they lack of a full detailed theoretical analysis.
We observe, however, that only for ultrarelativistic particles as neutrinos, changing helicity approximately means changing chirality.
What we intend to demonstrate in this paper is that, in the context of oscillation phenomena and
in the framework of a first quantized theory, the small differences between the concepts of chirality
and helicity, which had been interpreted as representing the same physical quantities for massless particles \cite{Oli90,Vol81,Bar96,Oli96,Kim93},
can be quantified for massive particles.
We concentrate our discussion, however, on the aspects concerning with chiral oscillations.
In the standard model flavor-changing interactions, neutrinos with positive
chirality are decoupled from the neutrino absorbing charged weak currents \cite{DeL98}.
Consequently, such positive chirality neutrinos become sterile with respect to weak interactions.
Independently of any external electromagnetic field, since neutrinos are detected essentially via V-A
weak charged currents, the chiral oscillation mechanism by itself
may even explain the ``missing'' LSND data.
Despite the experimental circumstances not
being completely favorable to such an interpretation, which should be an additional motivation
to our theoretical calculations, the quantum transitions that produces a final flavor eigenstate
corresponding to an active-sterile quantum mixing is perfectly acceptable
from the theoretical point of view. 
In fact, a formalism with Dirac wave packets \cite{Zub80,Ber04} leads to the study of chiral oscillations \cite{DeL98}.
Moreover, chiral coupled to flavor oscillations could lead to some small
modifications to the standard flavor conversion formula \cite{Ber05}.
Even focusing specifically on the chiral oscillation mechanism, 
we have also observed a generic interest in identifying the physical
meaning of variables which coexist with the interference between positive and negative frequency solutions of
Dirac equation \cite{Bra99,Rup00,Wan01,Bol04} in a Dirac wave packet treatment.
If we follow an almost identical line of reasoning as that applied for rapid oscillations
of the position, it is also possible to verify that the average value of the Dirac chiral operator $\gamma^{\5}$
also presents an oscillating behavior.

Our aim is thus just to clarify some conceptual relations between helicity and chirality by
constructing the dynamics of the processes of chiral oscillation and spin-flipping. 
The {\em differences} between the chirality ($\gamma^{\5}$) and the helcity ($h$) dynamics for a neutrino non-minimally coupling
with an external (electro)magnetic field are expressed in terms of the equation of the motion of the correspondent
operators $\gamma^{\5}$ and $h$.
To be more clear, in order to determine the relevant physical observables 
and confront the dynamics of chiral oscillations with the dynamics of spin flipping in the presence of an
external magnetic field, we report about a further class of static characteristics of neutrinos, namely, the (electro)magnetic
moment which appears in a Lagrangian with non-minimal coupling.
We shall demonstrate that the oscillating effects can be explained as an implication of the {\em zitterbewegung} phenomenon
that emerges when the Dirac equation solution is used for describing the time evolution of a wave packet \cite{Ber04}.
Due to this tenuous relation between {\em zitterbewegung} and chiral oscillations, the question to be 
answered in this manuscript concerns with the {\em immediate} description of chiral oscillations in terms of the {\em zitterbewegung} motion, i. e.
we shall demonstrate in the following that, in fact, chiral oscillations are coupled with the {\em zitterbewegung}
motion so that they cannot exist independently each order.

In order to make clear the confront of ideas involving
chirality and helicity, as an introductory purpose,
in section II, we recover such concepts and their common points at the same time 
we appoint their differences by
assuming a Lagrangian with minimal or non-minimal coupling .
In section III, we seek the {\em immediate} description of chiral oscillations in vacuum in
terms of the trembling motion ({\em zitterbewegung}) described by the velocity (Dirac) operator {\boldmath$\alpha$}.
We summarize this section with the interpretation of chiral oscillations as
position very rapid oscillation projections onto
a decomposed direction of the motion which is longitudinal to the momentum of the particle.
We draw our conclusions in section IV.

\section{Chirality and helicity}

Let us firstly consider the time evolution of a spin one-half free particle given by the Dirac equation
\begin{equation}
\left(i \gamma^\mu \partial_{\mu} - m \right)\psi\bb{x} = 0,
\label{00}
\end{equation}
with the plane wave solutions given by
$\psi\bb{x} = \psi_{_{\pl}}\bb{x} + \psi_{_{\mi}}\bb{x}$ where
{\small\begin{eqnarray}
&&\psi_{_{\pl}}\bb{x} = \exp{[-  i \, p \, x]} \, u\bb{p} ~~~~\mbox{for positive frequencies and}\nonumber\\
&&\psi_{_{\mi}}\bb{x} = \exp{[+ i \, p \, x]} \, v\bb{p} ~~~~\mbox{for negative frequencies},
\label{01}
\end{eqnarray}}\normalsize
$x = (t, \mbox{\boldmath$x$})$,  $p$ is the relativistic {\em quadrimomentum},  $p = (E, \mbox{\boldmath$p$})$, and the relativistic
energy is represented by $E^{\2} = m^{\2}+ \mbox{\boldmath$p$}^{\2}$.
The free propagating mass-eigenstate spinors are written as \cite{Pes95}
{\small\begin{eqnarray}
u_{\s}\bb{p} &=& \frac{\gamma^\mu p_\mu + m}{\left[2\,E\,(m+E)\right]^{\frac{1}{2}}}  \, u_{\s}\bb{m,\,0} = \left(\begin{array}{r} \left(\frac{E+m}{2E}\right)^{\foh} \eta_{\s}\\ \\ \frac{\mbox{\boldmath$\sigma$}.\mbox{\boldmath$p$}}{\left[2\,E\,(E+m)\right]^{\foh}} \eta_{\s} \end{array}\right),\nonumber\\
v_{\s}\bb{p} &=& \frac{-\gamma^\mu p_\mu + m}{\left[2\,E\,(m+E)\right]^{\foh}} \, v_{\s}\bb{m,\,0} = \left(\begin{array}{r} \frac{\mbox{\boldmath$\sigma$}.\mbox{\boldmath$p$}}{\left[2\,E\,(E+m)\right]^{\foh}} \eta_{\s}\\ \\\left(\frac{E+m}{2E}\right)^{\foh} \eta_{\s} \end{array}\right),~~~
\label{02}
\end{eqnarray}}\normalsize
where $\eta_{\1,\2} = \left(\begin{array}{c}1 \\ 0\end{array}\right),\, \left(\begin{array}{c}0 \\ 1\end{array}\right)$.
By considering the gamma matrices (in the Dirac representation),
\begin{equation}
\gamma^{\0} = \left(\begin{array}{rr} \mathbf{1} & 0 \\ 0  & -\mathbf{1}\end{array}\right),
~~~~~~~~ \gamma^i = \left(\begin{array}{cc} 0 & \sigma^i \\ - \sigma^i  &  0 \end{array}\right)
~~~~~~~~ \mbox{and} ~~~~~~~~ \gamma^{\5} = \left(\begin{array}{rr}  0  & \mathbf{1} \\ \mathbf{1} & 0 \end{array}\right),
\label{501}
\end{equation}
where $\sigma^{i}$ are the Pauli matrices, 
we can observe the following $\gamma^{\5}$ matrix properties,
\begin{eqnarray}
\left(\gamma^{\5}\right)^{\dagger} = \gamma^{\5},~~~~~~
\left(\gamma^{\5}\right)^{\2} = 1~~~~~~\mbox{and}~~~~~~
\left\{\gamma^{\5}, \gamma^{\mu}\right\}=0,
\label{502}
\end{eqnarray}
which lead to the
commuting relation $\left[\gamma^{\5}, S^{\mu\nu}\right] = 0$,
where $S^{\mu\nu}$ are the generators of the {\em continuous Lorentz transformations}.
Forming out of Dirac field bilinears, it is convenient to define the {\em vector}
and the {\em pseudo-vector} currents respectively by
\begin{equation}
j^{\mu}\bb{x} = \overline{\psi}\bb{x} \gamma^{\mu} \psi\bb{x} ~~~~~~~~\mbox{and}~~~~~~~~ j^{\mu \5}\bb{x} = \overline{\psi}\bb{x} \gamma^{\mu}\gamma^{\5} \psi\bb{x}.
\label{503}
\end{equation}
By assuming that $\psi\bb{x}$ satisfies the free Dirac equation, we can compute the divergence of this current densities,
\begin{eqnarray}
\partial_{\mu} j^{\mu}\bb{x} &=& (\partial_{\mu}\overline{\psi}\bb{x}) \gamma^{\mu} (\partial_{\mu}\psi\bb{x})
			   			  = (i\, m \, \overline{\psi}\bb{x})\psi + \overline{\psi}(-i\, m\, \psi\bb{x}) 
						  = 0
\label{504}
\end{eqnarray}
which leads to the conservation of $j^{\mu}\bb{x}$.
When we couple the Dirac field to the electromagnetic field, $j^{\mu}\bb{x}$ will become the electric current density.
In a similar way, we can compute
\begin{eqnarray}
\partial_{\mu} j^{\mu \5}\bb{x} &=& 2 \, i \, m \,\overline{\psi}\bb{x} \gamma^{\5} \psi\bb{x}
\label{505}.
\end{eqnarray}
If $m = 0$, this current, often called the {\em axial vector current}, is also conserved.
In general, it is useful to define the chiral {\em left}-$L$ (negative chiral)
and {\em right}-$R$ (positive chiral) currents through the combinations of the operator $\gamma^{\5}$ by
\begin{equation}
j^{\mu}_L \bb{x} = \overline{\psi}\bb{x} \gamma^{\mu} \frac{1 - \gamma^{\5}}{2} \psi\bb{x} ~~~~~~~~\mbox{and}~~~~~~~~ j^{\mu}_R\bb{x} = \overline{\psi}\bb{x} \gamma^{\mu} \frac{1 + \gamma^{\5}}{2} \psi\bb{x}.
\label{506}
\end{equation}
Just when $m = 0$, these are respectively the electric current densities of
{\em left} and {\em right-handed} particles and they are separately conserved.
The two currents $j^{\mu}\bb{x}$ and $j^{\mu \5}\bb{x}$
are the Noether currents corresponding to the two gauge transformations
respectively represented by
\begin{equation}
\psi\bb{x} \rightarrow \exp{[ i \,\alpha]} \psi\bb{x} 
\label{508}
\end{equation}
and
\begin{equation}
\psi\bb{x} \rightarrow \exp{[ i \,\alpha\, \gamma^{\5}]} \psi\bb{x} 
\label{509}.
\end{equation}
The former one is a symmetry of the Dirac Lagrangian.
The later one, called {\em chiral transformation}, is a
symmetry of the kinetic part of the Dirac Lagrangian \cite{Pes95}.
The Noether theorem confirms that an {\em axial vector current}
related to a {\em chiral transformation} is conserved only if $m = 0$.

In parallel, we can define the operator
$h = \frac{1}{2}\mbox{\boldmath$\Sigma \cdot \hat{p}$}$,  
so that, by conveniently choosing the two component spinors $\eta_{\s}$,
the spinors $u_{\s}\bb{p}$ and $v_{\s}\bb{p}$ become
eigenstates of $h$ with eigenvalues $h = + \frac{1}{2}$ for $s=1$ and $h = - \frac{1}{2}$ for $s=2$ ($\hbar = c =1$).
Although the free-particle plane wave solutions can always be chosen to be
eigenfunctions of $h$, it is not possible to find solutions which are
eigenfunctions of {\boldmath$\Sigma \cdot \hat{n}$} with an arbitrary unitary vector {\boldmath$\hat{n}$}.
It occurs because the operator {\boldmath$\Sigma \cdot \hat{n}$} does not commute with
the free particle Hamiltonian unless $\mbox{\boldmath$\hat{n}$} = \pm \mbox{\boldmath$\hat{p}$}$
or $\mbox{\boldmath$p$} = 0$.
The operator $h$ that can be diagonalized simultaneously with the free particle Hamiltonian is
called the {\em helicity} operator \cite{Sak87}.
It is exactly the particle spin projection onto the momentum direction.
The particle with $h = + \frac{1}{2}$ is called {\em right-handed} and that one with $h = - \frac{1}{2}$ is called {\em left-handed}.
The helicity of a massive particle depends on the reference frame,
since one can always {\em boost} to a frame in which its momentum is in the
opposite direction (but its spin is unchanged).
For a massless particle, which travels at speed of light, it is not possible to perform such a {\em boost}. 
In certain way, helicity and chirality quantum numbers carry a kind of complementary information. 
In vacuum, the chirality is invariant under a {\em continuous Lorentz transformation} but it is not constant in time if the particle is massive.
The helicity is constant in time but it is not Lorentz invariant.
A chiral eigenstate can always be a linear combination of two helicity eigenstate.

By turning back to the central point of our manuscript where we intend to compare the dynamic processes of chiral oscillation
and spin flipping, let us analyze the possibilities of observing them via Lagrangian interacting terms.
Initially, we can construct a gauge invariant Lagrangian in a theory with initially massless fermions.
We can couple a Dirac fermion to a gauge field by assigning the chiral-$L$ fields $\psi_{i_{L}}\bb{x}$
to one representation of a gauge group $G$ and assigning the chiral-$R$ fields $\psi_{i_{R}}\bb{x}$ to a different representation.
Thus we can write
\begin{equation}
\mathcal{L} = \overline{\psi}\bb{x} \left[\partial_\mu - i g A^a_\mu\bb{x}\, T^a \left(\frac{1 - \gamma^{\5}}{2}\right)\right] \psi\bb{x},
\label{511}
\end{equation}
where it is evident that only negative chiral fields couple with gauge boson fields $A^a_\mu\bb{x}$.
It is straightforward to verify that the Lagrangian (\ref{511}) is invariant to the
following infinitesimal ($\alpha^a\bb{x}\,$) local gauge transformation \cite{Pes95}
\begin{eqnarray}
\psi\bb{x}  & \rightarrow & \left[1 +  i \alpha^a_\mu\bb{x}\, T^a \left(\frac{1 - \gamma^{\5}}{2}\right)\right] \psi\bb{x}, \nonumber\\
A^a_\mu\bb{x}\, & \rightarrow & A^a_\mu\bb{x} + g^{-1} \partial_\mu \alpha^a\bb{x} + f^{abc}A^b_\mu\bb{x}\, \alpha^c\bb{x},
\label{512}
\end{eqnarray}
where the corresponding algebra properties are given by the commuting relations of operators $T^a$,
\begin{equation}
\left[T^a, T^b\right]= i f^{abc} T^c.
\label{513}
\end{equation}
Since the chiral-$R$ fields are free fields, we can even eliminate these fields and write a gauge invariant Lagrangian for purely chiral-$L$ fermions.
The idea of gauge fields that couple only to chiral-$L$ fermions plays a central role in the construction of  the theory of weak interactions.
To work out the general properties of chirally coupled fermions, it is useful to rewrite  their Lagrangian with one further transformation.
Let us take the charge conjugate operator $C = i \gamma^{\2} \gamma^{\0}$ in order to write a charge conjugate field $\psi^c$ as
\begin{equation}
\psi^c = C \bar{\psi}^{T} = i \gamma^{\2} \psi^{*}
\label{514}
\end{equation}
so that the conjugate chiral-$R$ component of a field $\psi$, which would be given by $\psi^c_L$,
transforms as a chiral-$L$ quantity under a continuous Lorentz transformation,
i. e.
\begin{equation}
\psi^c_L = (\psi^c)_L = i \left(\frac{1 - \gamma^{\5}}{2}\right) \gamma^{\2} \psi^{*} = i \gamma^{\2} \left(\frac{1 + \gamma^{\5}}{2}\right)  \psi^{*} = i \gamma^{\2} \psi_R^{*}, 
\label{515}
\end{equation}
and, in the same way, 
the conjugate chiral-$L$ component of a field $\psi$, which would be given by $\psi^c_R$,
transforms as a chiral-$R$ quantity,
i. e.
\begin{equation}
\psi^c_R  = i \gamma^{\2} \psi_L^{*}, 
\label{516}
\end{equation}
where, for simplicity, we have omitted the $x$ dependence.
This property is important for observing what occurs in a gauge model with {\em L-R} symmetry \cite{Moh75,Ple93}.
In this class of models, chiral-$R$ neutrinos appear as interacting particles 
coupling with vector gauge bosons described by the generators of the algebra
related to a symmetry which brings the unification group properties.
Meanwhile, due to the vectorial characteristic of the interactions which are mediated by 
vector gauge bosons, the possibility of a left to right (or vice-versa) chiral conversion
via gauge coupling is impossible.
As we can observe in the following, the probability amplitude for a vector coupling between fields with opposite chiral quantum numbers
is null,
\begin{eqnarray}
\overline{\psi}_L \,A_{\mu}\gamma^{\mu}\,\psi_L \equiv\overline{\psi_R} \,A_{\mu}\gamma^{\mu}\,\psi_L  &\propto&
\left((1 +\gamma^{\5})\psi\right)^{\dagger}\,\gamma^{\0}\,A_{\mu}\gamma^{\mu} (1 -\gamma^{\5})\psi\nonumber\\ 
&=&\psi^{\dagger} \,(1 +\gamma^{\5})\,\gamma^{\0}\,A_{\mu}\gamma^{\mu} (1 -\gamma^{\5})\psi\nonumber\\
&=& \psi^{\dagger} \,\gamma^{\0}\,A_{\mu}\gamma^{\mu} \,(1 + \gamma^{\5})(1 -\gamma^{\5})\psi = 0,
\label{516B}
\end{eqnarray}
which forbids the possibility of chiral changes (oscillations) in this way.
Alternatively, we cannot abandon the idea of spin-inversion which does {\em not}
correspond to chiral conversion but is a very well established mechanism which emerges via an
external vector field interaction.

\subsection{Chirality and helicity dynamics for massive neutrinos in the presence of an external magnetic field}

In the minimal standard model, neutrinos are massless and thus have no magnetic moment.
This is due to the fact that the magnetic moment $\mu$ is defined as a coefficient of the coupling
between a fermionic field $\psi\bb{x}$ and the electromagnetic field $F^{\mu\nu}\bb{x}$ given by
\begin{equation}
\frac{1}{2}\,\mu\, \overline{\psi}\bb{x}\, \sigma_{\mu\nu}F^{\mu\nu}\bb{x} \, \psi\bb{x} + h.c.
\end{equation}
that is an effective low-energy interaction, for masses and energies well below the $W^{\pm}$ mass
where $\sigma_{\mu\nu} = \frac{i}{2}[\gamma_{\mu},\gamma_{\nu}]$ and $F^{\mu\nu}= \partial^{\mu}A^{\nu}\bb{x} - \partial^{\nu}A^{\mu}\bb{x}$.
In a minimal extension of the standard model in which somehow neutrinos become massive Dirac particles,
leaving the rest of the model unchanged, the neutrino magnetic moment arises at the one-loop level,
as does the weak contribution to the anomalous magnetic moment of a charge lepton \cite{Zub80}.
The value of $\mu$ can be read off from general formulas for the electromagnetic vertex,
to one-loop order, of an arbitrary fermion ($\ell$).
To leading order in $m_{\ell}^{\2}/m_{W}^{\2}$, the results are independent of $m_{\ell}$
and of the mixing matrix so that
it turns out to be proportional to the neutrino mass,
\begin{equation} 
\mu = \frac{3\, e \,G}{8 \sqrt{2}\pi^{\2}} m = \frac{3\, m_e \,G}{4 \sqrt{2}\pi^{\2}}\, \mu_{\B} \, m_{\nu}
= 2.7 \times 10^{\mi \1\0}\,\mu_{\B}\,\frac{m_{\nu}}{m_N}
\end{equation} 
where $G$ is the Fermi constant and $m_{N}$ is the nucleon mass.
By principle, for $m_{\nu}\approx 1 \, eV$, the magnetic moment introduced by this formula is exceedingly small
to be detected or to affect astrophysical or physical processes.
 
We have not discriminated the flavor/mass mixing elements in the coupling described above because we are initially interested just in
the physical observable dynamics ruled by the effective Hamiltonian 
\begin{eqnarray} 
\mathit{H} &=& \mbox{\boldmath$\alpha$}\cdot \mbox{\boldmath$p$} + \beta \left[m - \left(\frac{1}{2}\,\mu \,\sigma_{\mu\nu}F^{\mu\nu}\bb{x} + h.c.\right)\right]
\nonumber\\
 		   &=& 
=\mbox{\boldmath$\alpha$}\cdot \mbox{\boldmath$p$} + \beta \left[m - Re(\mu) \mbox{\boldmath$\Sigma$}\cdot \mbox{\boldmath$B$}\bb{x} +Im(\mu) \mbox{\boldmath$\alpha$}\cdot \mbox{\boldmath$E$}\bb{x}\right] 
\label{08},~~
\end{eqnarray} 
where $\mbox{\boldmath$\alpha$} = \sum_{\k \ig \1}^{\3} \alpha_{\k}\hat{\kk} = \sum_{\k \ig \1}^{\3} \gamma_{\0}\gamma_{\k}\hat{\kk}$,
$\beta = \gamma_{\0}$, $\mbox{\boldmath$B$}\bb{x}$ and $\mbox{\boldmath$E$}\bb{x}$ are respectively the magnetic and electric fields.
We have also assumed that $\mathit{H}$ represents the time evolution operator
for times subsequent to the creation of a spinor $w$ $(t = 0)$.
The real (imaginary) part of $\mu$ represents the magnetic (electric) dipole moment of the mass eigenstate represented by
$\psi\bb{x}$.
But it can be demonstrated \cite{Kim93} that, for Dirac neutrinos, the electric moment must vanish unless $CP$ is violated,
and, for Majorana neutrinos, the electric moment vanishes if $CPT$ invariance is assumed.

Turning back to our main assertions, 
the problem to be solved concerns with the {\em immediate} description of chiral oscillations and spin flipping 
in terms of the Hamiltonian (\ref{08}) dynamics by recurring to the Heisenberg equation of motion \cite{Sak87},
\begin{equation} 
\frac{d~}{dt}\langle\mathcal{O}\rangle = i \langle\left[ \mathit{H} ,\mathcal{O} \right]\rangle
+ \langle\frac{\partial \mathcal{O}}{\partial t}\rangle, 
\label{09}
\end{equation} 
which show us how to easily determine 
whether or not a given observable $\mathcal{O}$ is a constant of the motion.
By means of very simple calculations,
we notice, for instance, that the free propagating particle momentum is not a conserved quantity for the above Hamiltonian (\ref{08}),
\begin{equation} 
\frac{d~}{dt}\langle\mbox{\boldmath$p$}\rangle \,=\, Re\bb{\mu} \langle \beta\, \mbox{\boldmath$\nabla$} \left(\mbox{\boldmath$\Sigma$}\cdot \mbox{\boldmath$B$}\bb{x}\right)\rangle.
\label{100}
\end{equation} 
In the same way, the particle velocity given by
\begin{equation} 
\frac{d~}{dt}\langle\mbox{\boldmath$x$}\rangle \,=\, i\langle\left[ \mathit{H} , \mbox{\boldmath$x$}\right]\rangle\,=\,\langle\mbox{\boldmath$\alpha$}\rangle 
\label{11}
\end{equation}
comes out as a non-null value.
After solving the $\mbox{\boldmath$x$}\bb{t}$($\mbox{\boldmath$\alpha$}\bb{t}$) differential equation,
it is possible to observe that the fermionic particle
executes very rapid oscillations in addition to a uniform motion.
This quivering motion is known as {\em zitterbewegung} \cite{Sak87}.
By following an almost identical line of reasoning as that applied for rapid oscillations
of the position, it is possible to verify that the average value of the Dirac chiral operator $\gamma^{\5}$
also presents an oscillating behavior.
By newly recurring to the Heisenberg equation (\ref{09}), it is possible to obtain
the chirality and the helicity dynamics respectively as
\begin{eqnarray} 
\frac{d~}{dt}\langle \gamma^5 \rangle &=& 2\,i \langle\gamma_{\0}\,\gamma_{\5} \left[m - Re(\mu) \mbox{\boldmath$\Sigma$}\cdot \mbox{\boldmath$B$}\bb{x}\right]\rangle
\label{12}
\end{eqnarray} 
and
\begin{eqnarray} 
\frac{d~}{dt}\langle h \rangle &=& \frac{1}{2}\,Re(\mu)\langle \gamma_{\0} \left[(\mbox{\boldmath$\Sigma$}\cdot\mbox{\boldmath$\nabla$})(\mbox{\boldmath$\Sigma$}\cdot\mbox{\boldmath$B$}\bb{x})
+ 2(\mbox{\boldmath$\Sigma$}\times \mbox{\boldmath$B$}\bb{x})\cdot \mbox{\boldmath$p$}\right]\rangle
\label{13}
\end{eqnarray} 
where we have alternatively redefined the particle helicity as 
the projection of the spin angular momentum onto the vector momentum,
$h = \frac{1}{2}\mbox{\boldmath$\Sigma$}\cdot{\mbox{\boldmath$p$}}$ (with
${\mbox{\boldmath$p$}}$ in place of $\hat{\mbox{\boldmath$p$}}$).
Therefore, we remark that, if a neutrino has an intrinsic magnetic moment and passes through a region filled by
an external magnetic field, the neutrino helicity can flip in a completely different way from how
chiral oscillations take place.
The Eqs.~(\ref{12}-\ref{13}) can be reduced to the non-interacting case by setting $\mbox{\boldmath$B$}\bb{x} = 0$.
In this case,
\begin{equation} 
\frac{d~}{dt}\langle h \rangle \,=\, i\langle\left[ \mathit{H} , h\right]\rangle\,=\,- \langle(\mbox{\boldmath$\alpha$} \times \mbox{\boldmath$p$})\cdot\hat{\mbox{\boldmath$p$}}\rangle \,=\ 0
\label{16}
\end{equation} 
and, in the same way, the chiral operator $\gamma^{\5}$ is {\em not} a constant of the motion \cite{DeL98} since
\begin{equation} 
\frac{d~}{dt}\langle\gamma^{\5}\rangle \,=\, i\langle\left[ \mathit{H} , \gamma^{\5}\right]\rangle\,=\,2 \,i\,m \langle\gamma^{\0}\gamma^{\5}\rangle
\label{17}.
\end{equation} 
The effective value of Eq.~(\ref{13}) appears only when both positive and negative frequencies are taken into account
to compose a Dirac wave packet, i. e. 
the non-null averaged value of $\langle\gamma_{\0}\gamma_{\5}\rangle$ is revealed by the interference
between Dirac equation solutions with opposite sign frequencies.
In fact, in order to quantify neutrino chiral oscillations in vacuum, we could report to the Dirac wave packet formalism
\cite{Ber05} where the general procedure consists in writing the general (complete) Dirac equation (wave packet) solution of Eq.~(\ref{01}) as
\begin{eqnarray}
\fl\psi\bb{t, \mbox{\boldmath$x$}}
= \int\hspace{-0.1 cm} \frac{d^{\3}\hspace{-0.1cm}\mbox{\boldmath$p$}}{(2\pi)^{\3}}
\sum_{\s \ig \1,\2}\{b_{\s}\bb{p}u_{\s}\bb{p}\, \exp{[- i\,E\,t]}
+ d^*_{\s}\bb{\tilde{p}}v_{\s}\bb{\tilde{p}}\, \exp{[+i\,E\,t]}\}
\exp{[i \, \mbox{\boldmath$p$} \cdot \mbox{\boldmath$x$}]}~~
\label{03}
\end{eqnarray}
where $\overline{u}\bb{p}$ (or $\overline{v}\bb{p}$) is defined by
$\overline{u}\bb{p} = u^{\dagger}\bb{x}\gamma^{\0}$ (or $\overline{v}\bb{p} = v^{\dagger}\bb{p}\gamma^{\0}$)
and $\tilde{p} = \bb{E, -\mbox{\boldmath$p$}}$.
By fixing the initial condition over $\psi\bb{0,\mathbf{x}}$
as the Fourier transform of the weight function 
\begin{equation}
\varphi\bb{\mbox{\boldmath$p$}-\mbox{\boldmath$p$}_{\0}}\,w \,=\,
\sum_{\s \ig \1,\2}\{b_{\s}\bb{p}u_{\s}\bb{p} + d^*_{\s}\bb{\tilde{p}}v_{\s}\bb{\tilde{p}}\}
\label{002}
\end{equation}
we get
\begin{equation}
\psi\bb{0, \mbox{\boldmath$x$}}
= \int\hspace{-0.1 cm} \frac{d^{\3}\hspace{-0.1cm}\mbox{\boldmath$p$}}{(2\pi)^{\3}}
\varphi\bb{\mbox{\boldmath$p$}-\mbox{\boldmath$p$}_{\0}}\exp{[i \, \mbox{\boldmath$p$} \cdot \mbox{\boldmath$x$}]}
\,w
\label{003}
\end{equation}
where $w$ is some fixed normalized spinor.
The coefficients $b_{\s}\bb{p}$ and $d^*_{\s}\bb{\tilde{p}}$ can thus be calculated by
using the orthogonality properties of Dirac spinors.
These coefficients carry an important physical information.
For {\em any} initial state $\psi\bb{0,\mathbf{x}}$ given by Eq.~(\ref{003}),
the negative frequency solution coefficient
$d^*_{\s}\bb{\tilde{p}}$ necessarily provides a non-null contribution to the time evolving wave packet.
This obliges us to take the complete set of Dirac equation solutions to construct the wave packet.
Only if we consider a momentum distribution given by a delta function (plane wave limit) and suppose an initial spinor $w$ being a positive energy mass-eigenstate with momentum
$\mbox{\boldmath$p$}$, the contribution due to the negative frequency solutions 
$d^*_{\s}\bb{\tilde{p}}$ will become null.

By following a similar procedure, we could also calculate the
averaged values which appear in Eqs.~(\ref{12}-\ref{13}).
But we would have to be careful when calculating the exact Hamiltonian eigenstates and eigenvalues in order to construct
an accurate superposition of states which necessarily respect the orthonormality relations and represent a complete set of solutions.
This point does not necessarily correspond to a problem with trivial solutions, however, 
in a first analysis, as a suggestion, one could ignore the space-temporal analytic characteristics
of the external magnetic field $\mbox{\boldmath$B$}$ and
assume the simplifying hypothesis where $\mbox{\boldmath$B$}$ is a uniform magnetic field so that
\begin{eqnarray} 
\frac{d~}{dt}\langle \gamma^5 \rangle &=& 2\,i \langle\gamma_{\0}\,\gamma_{\5} \left[m - Re(\mu) \mbox{\boldmath$\Sigma$}\cdot \mbox{\boldmath$B$}\right]\rangle
\label{12A}.
\end{eqnarray} 
and
\begin{eqnarray} 
\frac{d~}{dt}\langle h \rangle &=& Re(\mu)\langle \gamma_{\0} 
(\mbox{\boldmath$\Sigma$}\times \mbox{\boldmath$B$})\cdot \mbox{\boldmath$p$}
\rangle
\label{13A}.
\end{eqnarray} 
and thus proceed with the calculations.

Effectively the above discussion represents the first step for accurately deriving the expression for the neutrino
spin-flipping in magnetic field which can be related to chiral oscillations in the limit of a massless particle (ultra-relativistic limit).
By correctly differing the concepts of helicity and chirality, we can determine the origin and the influence of chiral oscillations and spin-flipping
in the complete flavor conversion formula.

In some previous manuscripts \cite{Ber05A,Ber04,Ber05}
we have confirmed that the {\em fermionic} character of the particles modify the standard oscillation
probability which is previously obtained by implicitly assuming a {\em scalar} nature of
the mass-eigenstates.
Strictly speaking, we have obtained the term of very high oscillation frequency depending on the sum of energies
in the new oscillation probability formula which, in case of Dirac wave-packets, represents
modifications that introduce correction factors which, under the current experimental point of view,
are not effective in the UR limit of propagating neutrinos in vacuum but which deserves, at least, a careful investigation
for neutrino moving in the background matter where spin/chiral effects become more relevant \cite{Kim93,MSW,Guz03}.
The physical consequences in environments such as supernova can be theoretically studied \cite{Oli99}.
For instance, it was observed that neutrinos propagating in matter achieve an effective electromagnetic vertex
which affects the flavor conversion process in a framework where preserving chirality can be established \cite{Oli96}.

Finally, just as a remark, it is also to be noted that in this kind of analysis we have to assume that neutrinos are
Dirac particles, thus making the positive-chiral component sterile. 
If neutrinos are Majorana particles, they cannot have a magnetic moment, obviating the spin-flipping via magnetic field interactions
but still allowing the (vacuum) chiral conversion possibility via very rapid oscillations ({\em zitterbewegung}) 
as we shall present in the next section.

\section{\label{sec4}Conclusions}

In this brief study, we have discussed the concepts of chirality and helicity quantum numbers which appear
when we construct the interacting Lagrangian of the electroweak theory.
The effects of chiral oscillations were explained as a consequence of the
{\em zitterbewegung} effect which emerges when Dirac equation solutions are used for describing the
space-temporal evolution of a wave packet.
In particular, we seek the {\em immediate} description of chiral oscillations in
terms of the trembling motion described by the velocity (Dirac) operator {\boldmath$\alpha$}.
By taking into account the complete
set of Dirac equation solutions which results in a free propagating Dirac wave packet composed by
positive and negative frequency components, we report about the well-established 
{\em zitterbewegung} results and indicate how chiral oscillations can be given in terms of the already
know quantum oscillating variables.    
We have also interpreted chiral oscillations as position very rapid oscillation projections onto
longitudinal (momentum) decomposed direction.
By comparing the dynamic characteristics of chiral oscillations with
the process of spin flipping which is related to the quantum number of helicity,
we also have noticed that there is not a constraint between both phenomena in th sense that they are
described by different dynamic evolution equations.

The clear understanding of the independent characteristics of
the process of chiral oscillations and spin-flipping is important
for several extensions of the quantum oscillation phenomena.
Once we have assumed the electroweak interactions at the source and detector are ({\em left}) chiral
$\left(\overline{\psi} \gamma^{\mu}(1 - \gamma^{\5})\psi W_{\mu}\right)$
only the component with negative chirality contributes to the propagation.
In this case, chiral coupled to flavor oscillations could lead to some small
modifications to the standard flavor conversion formula \cite{Ber05}.
We remind that, in the standard treatment of
vacuum neutrino oscillations, the use of scalar mass-eigenstate wave packets made up
exclusively of positive frequency plane-wave solutions is usually implicitly
assumed. Although the standard oscillation formula could give the correct
result when {\em properly} interpreted, a satisfactory description
of fermionic (spin one-half) particles requires the use of the
Dirac equation as evolution equation for the mass-eigenstates.
Consequently, the the spinorial form and
the interference between positive and negative frequency
components of the mass-eigenstate wave packets could lead to the possibility
of chiral coupled with flavor oscillations which effectively introduce some small
modifications to the {\em standard} flavor conversion formula \cite{Ber04} when it
is applied to the study of non-relativistic neutrinos.

Our future perspectives concern with deriving the flavor coupled with chiral conversion expressions for neutrinos
moving in the background matter by supposing that the magnitude of some experimentally (implicitly) observable
matter effect could be quantified (and eventually detected).
We know, however, the necessity of a more sophisticated approach
is understood. In fact, the derivation of the oscillation formula should resort to field-theoretical methods
which, meanwhile, are not very popular.
They are thought to be very complicated and the existing quantum field computations of the
oscillation formula do not agree in all respects \cite{Beu03}.
The  Blasone and Vitiello (BV) model \cite{Bla95,Bla03} to neutrino/particle mixing and oscillations
seems to be the most distinguished trying to this aim.
They have attempt to define a Fock space of weak eigenstates and
to derive a nonperturbative oscillation formula.
Flavor creation and annihilation operators, satisfying canonical
(anti)comutation relations, are defined by means of Bogoliubov transformations.
As a result, new oscillation formulas are obtained for fermions and
bosons, with the oscillation frequency depending not only on the
difference but also on the sum of the energies of the different mass eigenstates .
In the particular framework of Dirac wave-packets, before aggregating the chiral oscillation
effects, the flavor conversion formula can be reproduced \cite{Ber04}
with the same mathematical structure as those obtained in the BV model \cite{Bla95,Bla03}.
Moreover each new effect present in the oscillation formula can be separately quantified.
The quantization of chiral oscillations in the framework of a quantum field theory can
be eventually cogitated as a next step.
  
\subsection*{Acknowledgments}

The author thanks FAPESP (PD 04/13770-0) for Financial Support.

\end{document}